# CONTROLLABLE SEQUENCE-TO-SEQUENCE NEURAL TTS WITH LPCNET BACKEND FOR REAL-TIME SPEECH SYNTHESIS ON CPU

*Slava Shechtman[1], Carmel Rabinovitz[1], Alex Sorin, Zvi Kons, Ron Hoory*

IBM Research, Haifa, Israel

slava@il.ibm.com

## ABSTRACT

State-of-the-art sequence-to-sequence acoustic networks, that convert a phonetic sequence to a sequence of spectral features with no explicit prosody prediction, generate speech with close to natural quality, when cascaded with neural vocoders, such as Wavenet. However, the combined system is typically too heavy for real-time speech synthesis on a CPU.

In this work we present a sequence-to-sequence acoustic network combined with lightweight LPCNet neural vocoder, designed for real-time speech synthesis on a CPU. In addition, the system allows sentence-level pace and expressivity control at inference time.

We demonstrate that the proposed system can synthesize high quality 22 kHz speech in real-time on a general-purpose CPU. In terms of MOS score degradation relative to PCM, the system attained as low as 6.1-6.5% for quality and 6.3-7.0% for expressiveness, reaching equivalent or better quality when compared to a similar system with a Wavenet vocoder backend.

***Index Terms***—sequence to sequence, neural TTS, neural vocoder, LPCNet, expressive TTS, speech prosody

## 1. INTRODUCTION

Modern sequence-to-sequence neural TTS systems can generate speech with close-to-natural speech quality [1][2]. Usually, such systems obtain an input linguistic sequence (e.g. phonemes, enriched with other text-based features, such as lexical stress, phrasing, word embedding, etc.) and output a speech acoustic sequence, represented by frame-wise spectral parameters (e.g. mel-spectrum), from which a waveform can be generated using neural vocoders [10][4] or by low-quality (but much faster) signal processing algorithms [5][6]. Such systems deploy a convolutional and/or recurrent multi-layer linguistic encoder followed by a multi-layer auto-regressive attentive acoustic decoder [1][2]. They generate speech prosody (i.e. phone durations, pitch and loudness trajectories) implicitly, thus their prosody control is not a straightforward task [7]. However, they tend to generate speech with better quality, compared to modular neural TTS systems, that comprise several non-autoregressive multi-layer networks with explicit and easily controllable prosody generation [8][9].

In our previous work [7] a sequence-to-sequence TTS system with prosody modification capabilities was introduced. The system deploys Tacotron2 encoder-decoder architecture [2], with *prosody info* controls [7] of sentence-wise speaking pace and expressiveness. This system employs an *augmented attention* mechanism [7] designed to improve the system robustness when applying the prosody controls at inference time. The waveform generation is performed by a Wavenet[10] based neural vocoder [2][3] operating on mel-spectrogram, as originally proposed in [2]. However, the computational complexity of this neural vocoder is too high for real-time TTS systems running on a general-purpose CPU.

Recently, an efficient neural vocoder, called LPCNet, was introduced [4] and successfully deployed in a modular neural TTS system [11]. The LPCNet inference runs faster than real-time on a single CPU while producing a high-quality speech output. LPCNet obtains as inputs cepstral coefficients representing a spectral envelope, pitch and voicing features.

In this paper we present a modified version of the controllable sequence-to-sequence TTS with *prosody info* [7], suitable for the waveform generation by LPCNet neural vocoder. We demonstrate that the proposed system can generate 22 kHz speech faster than real-time on a general-purpose CPU, while providing similar or better quality as compared to the previously reported system [7] with Wavenet backend.

The paper is structured as follows. In Section 2 the controllable sequence-to-sequence TTS with *prosody info* [7] is reviewed and a simplified *augmented attention* mechanism used in the current system is presented. In Section 3 various options for LPCNet input features are discussed. The LPCNet feature prediction by the sequence-to-sequence TTS system is described in Section 4. Finally, Section 5 presents experimental results demonstrating that the proposed system with LPCNet neural vocoder fits in real-time on a general-purpose CPU and provides equivalent or better perceived speech quality as compared to the similar system [7] with the Wavenet vocoder backend.

---

[1] Authors with equal contribution

## 2. ACOUSTIC SEQUENCE PREDICTION NETWORK

### 2.1. Overview

The sequence-to-sequence acoustic feature prediction module (See Figure 1) extends the Tacotron2 [2] architecture with the *augmented attention* mechanism and the *prosody info* control [7] at inference time.

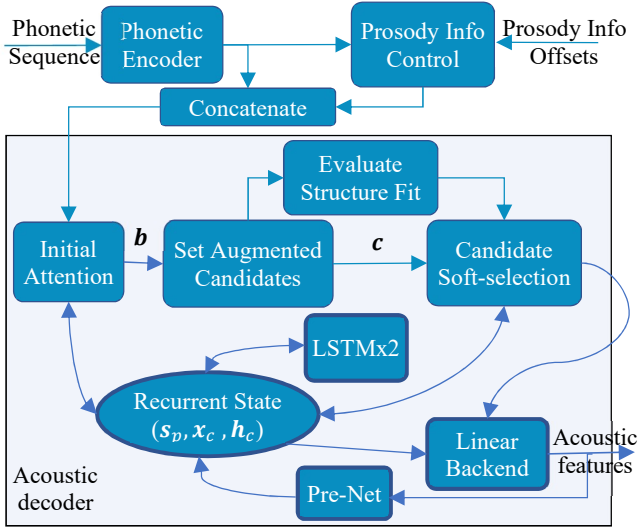

**Figure 1:** Acoustic sequence prediction network with prosody info control and augmented attention. (The refinement post-net [2] is omitted for the simplicity.)

The Tacotron2 [2] architecture comprises a convolutional encoder with a terminal recurrent layer (bidirectional [12] LSTM [13]), converting the symbolic sequence to the sequence of their latent representations, cascaded with the autoregressive attentive decoder that expands the encoded symbolic sequence to a sequence of constant frame (12.5 ms) mel-spectral feature vectors (of 80 coefficients).

The Tacotron2 decoder [2] predicts one frame at a time from the pre-net-processed previous frame $s_p$ conditioned on the input context vector $x_c$ generated by the attention module. The decoder generates its hidden state vector $h_c$ with the two-layered stacked LSTM network. The hidden state vector $h_c$ combined with the input context vector $x_c$ is fed to the two parallel fully connected (FC) linear layers (referred to as Linear Backend) producing the current mel-spectrum and the binary end-of-sequence flag correspondingly. The latter layer is terminated with the sigmoid non-linearity. At the end, there is also an optional convolutional post-net [2] (omitted in Figure 2 for the simplicity) that refines the whole utterance mel-spectrogram to improve its fidelity. In addition, the autoregressive decoder features teacher-forcing training with *double feedback* [7] and the *differential spectral MSE loss* [7].

The system employs the optional prosody control mechanism, based on *prosody info* observations (log-duration and log-pitch span), as proposed in our previous work [7]. The utterance-level prosody controls (i.e. the *prosody info* normalized offsets) can be deliberately added at inference time to control the pace and expressiveness in the synthesized speech. When properly tuned (per voice and desired speaking style), the prosody info offsets can improve the perceived quality [7], so we preserve this mechanism in the current work.

The attention mechanism, utilized in the decoder, is composed of the primary content-based and location-based attention proposed in Tacotron2 [2] followed by the structure-preserving *augmented attention* mechanism [7] designed to improve the system robustness when applying the prosody control at inference time. In this work we used a simplified formulation of the candidate soft selection operation within the augmented attention [7] (See also Figure 1) as elaborated below.

### 2.2. Augmented attention mechanism

The attention mechanism is essential in the decoder module for automatic alignment between the decoder input and output sequences. At each output time step $t \epsilon \mathbb{Z}$, $0 \leq t < T$, corresponding to the output frame sequence index $t$, the vector of attention weights is produced (i.e. *alignment vector* $a_t[n], 0 \leq n < N$) determining the relative attention (i.e. the *alignment* probabilities) of the $t$-th output to the $n$-th element of the input sequence, where $n \epsilon \mathbb{Z}$, $0 \leq n < N$. Once the alignment vector is determined, the input context vector $x_c$ is obtained as a convex linear combination of the decoder input sequence vectors with the alignment vector components serving as their weights. The input context vector is stored as a part of the decoder state and fed to the next decoder stages to finalize the output sequence prediction. This process is repeated until the end-of-sequence symbol is generated.

Let $b_t[n]$ be the initial alignment vector as evaluated by the initial attention module and $a_t$ be a final alignment vector at output time step $t$. As we originally proposed in [7], we create an alignment vector candidate set $\{c_k\}_{k=0}^2$ by adding the previous alignment vector $b_{t-1}[n]$ together with its shifted version $b_{t-1}[n-1]$ to the current initial alignment $b_t$ at the current time step $t$:

$$\{c_k\}_{k=0}^2 = \{b_t, b_{t-1}[n], b_{t-1}[n-1]\} \quad (1)$$

Having this set of candidates, we evaluate a scalar structure fit metric $f(c_k)$ that assesses the unimodality and the peak sharpness of each alignment vector candidate $c_k$. The metric is differentiable and confined in [0,1] interval (please refer to [7] for the exact formula used).

In this work we utilize a single stage candidate soft-selection. Let $\widehat{log}(x)$ be a confined logarithm operator, e.g. $\widehat{log}(x) = \max(\log(x), -50)$. Then for each attention vector candidate we evaluate its structure fit penalty $p_k = \widehat{log}(f(c_k))$.

We utilize the vector of structure fit penalties $p$ in a single-stage candidate soft-selection as follows. Let $FC(S)$ be an FC layer fed with the concatenated decoder state variables $S =$

$(s_p, x_c, h_c)$. Then the soft-selection weights $\alpha = \{\alpha_k\}_{k=0}^{2}$ are predicted by:

$$\alpha = softmax(FC(S) + p) \qquad (2)$$

Apparently, the structure fit penalty would shift the selection weight of an ill-formed candidate to zero, thus reducing its influence on the final alignment vector $a$ that is obtained by the soft-selection operation $a = \sum_k \alpha_k c_k$.

In our experiments the proposed single-stage candidate selection provides about the same synthesized speech quality as the two-stage soft-selection used in [7] while the former is simpler and directly generalize to any number of attention vector candidates.

## 3. LPCNET NEURAL VOCODER

### 3.1. Features

The LPCNet [4] vocoder, like other neural vocoders, generates speech samples from a sequence of equidistant-in-time acoustic feature vectors. For 16 kHz speech each vector contains 18 cepstral coefficients, one pitch parameter and one voicing (pitch correlation) parameter [4]. Unlike other waveform generative models, LPCNet predicts the Linear Prediction (LP) residual (the vocal source signal) and then applies to it an LP filter calculated from the cepstrum. To suit the sequence-to-sequence acoustic feature generation for the 22 kHz speech, we used a frame hop of 256 samples and an analysis window of 512 samples with 512-point FFT for the cepstrum calculation. We added two more frequency channels: 8000 – 9500 Hz and 9500 – 11025 Hz yielding 20 cesptral coefficients per frame (i.e. 22 LPCNet features altogether).

For the pitch estimation, we used our proprietary pitch estimator [14] with log-pitch contour linearly interpolated over unvoiced regions. Then the log-pitch values that exceed a predefined range of $[log60, log360]$ are clipped. Finally, the log-pitch values are quantized to 256 levels.

As for the pitch correlation, we used the maximal normalized autocorrelation value of the audio signal within the analysis window. The maximal value is calculated within the lag range corresponding to the pitch interval $[60\,Hz, 360\,Hz]$. Negative values are replaced by $0$.

### 3.2. Experimenting with alternative features

At the early stages of our work we tried to develop LPCNet integration methods that keep the acoustic feature prediction intact, so that the 80-channel mel-spectrum outputs from the sequence-to-sequence acoustic prediction network are directly fed to an LPCNet model trained on such mel-spectra. In that case the pitch and the voicing features are not explicitly presented to the neural vocoder but rather contained implicitly in the input mel-spectrum. To that end, we slightly modified the LPCNet model architecture removing the pitch embedding layer [4] which became irrelevant.

In addition, we developed an algorithm for the LP parameters estimation from the mel-spectrum. The major steps of the algorithm were smoothing and mel-to-linear resampling of the input mel-spectrum for a spectral envelop estimation.

However, subjective listening evaluation results revealed that although this method provided quite good results for a female voice, the quality dropped significantly for a male voice. In our opinion, it can be explained by the fact that the pitch information in mel-spectrum is more evident for high-pitched speech (female) than for the low-pitched speech (male). At that stage we decided to switch to a direct prediction of LPCNet parameters by the sequence-to-sequence acoustic feature predictor.

## 4. LPCNET FEATURE PREDICTION

### 4.1. Architecture consideration

The autoregressive attentive decoder is known to be quite sensitive to the type of the predicted spectral features. Indeed, the direct substitution of the mel-spectra by LPCNet parameters in the acoustic feature prediction network resulted in significant quality and intelligibility degradation.

Instead, we extended the linear backend only (see Figure 1) with an additional 2-layered feed forward (FF) FC network (hidden sizes of 512, 256 nodes, with the intermediate *tanh* non-linearity) to predict 22 LPCNet features from the hidden state vector $h_c$ concatenated with the input context vector $x_c$.

First, we tried to train the LPCNet prediction FF network separately after the whole mel-spectrogram prediction network had been trained. Note that in this case the prosody of the original mel-spectrogram prediction network is preserved. This approach provides a decent synthesized speech quality. However, it was inferior compared to the Wavenet vocoder baseline.

We also experimented with LPCNet parameters' feedback (either all, or pitch plus voicing parameters only) in addition to the regular mel-spectral feedback, however, the perceived quality did not improve either.

Eventually, we came up with joint multi-task learning process, predicting both mel-spectrum and LPCNet features, with the convolutional post-net [2] applied to the LPCNet features. As the LPCNet parameters are heterogenous, we applied two parallel 5-layered convolutional post-nets, the first is for the 20 cepstral coefficients (512 filters each layer, kernel size of 5), and the second is for the log-pitch and pitch correlation (64 filters each layer, kernel size of 5)

### 4.2. Training loss

The joint spectral training loss combines the absolute MSE losses for mel-spectrum and LPCNet parameters prior to the post-nets and the absolute and the differential MSE losses for LPCNet parameters after the post-nets.

Let $y_{M,t}$, $y_{L,t}$ be the predicted vectors at time *t* before the post-nets, for mel-spectrum and LPCNet features respectively; $z_{L,t}$ be the final predicted LPCNet feature vector at time *t*; and $q_{M,t}$, $q_{L,t}$ be the mel-spectrum and the LPCNet feature targets at time *t* respectively. Then the combined spectral loss is given by:

$$Loss_{spc} = MSE(y_{M,t}, q_{M,t}) + 0.8\, MSE(y_{L,t}, q_{L,t}) +$$
$$0.4 MSE(z_{L,t}, q_{L,t}) + 0.4 MSE(z_{L,t} - z_{L,(t-1)}, q_{L,t} - q_{L,(t-1)}) \quad (3)$$

The above combined spectral loss is added to the binary end-of-sequence flag cross-entropy loss to yield the total training loss.

## 5. EXPERIMENTS

Measuring the performance of the proposed 22 kHz-speech LPCNet feature sequence generation on Intel Xeon CPU (2.7 GHz) we obtained 0.43 real time factor (RTF) for single-threaded and 0.27 RTF for double-threaded *pytorch*[2] (v.1.0.0) runs. The similar measurements of the LPCNet vocoder performance [11] yields 0.25 RTF. Hence the whole system performs synthesis faster than real-time.

For subjective speech assessment experiments, we trained the system on professionally recorded 22kHz native US English male (13 hours long) and female (22 hours long) proprietary voice corpora.

We have conducted one formal MOS listening test per voice. In each test we assessed several systems' quality and expressiveness. We used each system to synthesize a set of 40 held-out sentences and evaluated them together with the original held-out recordings of the same voice. A subset of samples for the experiments described below is accessible online[3].

The tests were performed using the Amazon Mechanical Turk platform with 98–126 anonymous subjects, so that each sentence was evaluated by 25 distinct subjects. More details on our listening test procedure can be found in [7]. The systems that participated in each test included:

- "PCM": held out speech recordings at 22kHz;
- "mod-wrld": modular neural TTS with WORLD vocoder [6] at 22 kHz described in [9];
- "mod-lpc": Modular neural TTS with LPCNet backend [11] at 22 kHz;
- "seq2wv(*a,b*)": the controllable seq2seq system with Wavenet backend with the prosody info offsets *a* and *b* for pace and expressiveness components respectively that performed the best in our previous experiments [7];
- "seq2lpc(*a,b*)": the proposed controllable seq2seq system with the LPCNet backend, the single stage augmented attention mechanism and the prosody info offsets *a* and *b* for pace and expressiveness components respectively.

In Tables 2 and 3 the speech quality and expressiveness evaluation results are presented for the female and the male voice respectively including the average score±95% and the relative difference to PCM. The significance analysis for the results in Table 2 revealed that most of cross-system expressiveness differences are statistically significant, except for the difference between *seq2wv*(0.15,0.6) and *seq2lpc*(0.15,0.6). In terms of quality, *mod-lpc*, *seq2wv*(0.15,0.6) and *seq2lpc*(0.15,0.6) performed the same, while *seq2wv*(0.0,0.8) performed the best. Thus, for the female voice the proposed system with LPCNet backend was able to significantly improve both perceived quality and expressiveness compared to all the other systems. The significance analysis for the male voice (Table 3) revealed that both of the proposed systems (*seq2lpc(a,b)*) are equivalent in terms of perceived expressiveness and slightly worse than the best baseline system with Wavenet backend (*seq2wv*(0.2,0.8)). In terms of quality, both of the proposed systems (*seq2lpc(a,b)*) are equivalent to the best baseline system (*seq2wv*(0.2,0.8)). Thus, for the male voice the proposed system with LPCNet backend was able to achieve the statistically equivalent (*p*=0.06) naturalness, but slightly reduced expressiveness (*p*=0.035) compared to the best baseline system with the Wavenet backend. However, the proposed system significantly improved the quality and expressiveness compared to the modular TTS with LPCNet [10][11]. In terms of relative gap to PCM, the system attained as small gap as 6.1-6.5% for the quality and 6.3-7.0% for the expressiveness.

**Table 1.** *MOS Evaluation for US English female voice ($\mu\pm95\%$; $100(\mu - \mu_{PCM})/\mu_{PCM}$)*

| MOS | PCM | mod-wrld [9] | mod-lpc [11] | seq2wv (0.15,0.6) | seq2lpc (0.15,0.6) | seq2lpc (0.0,0.8) |
|---|---|---|---|---|---|---|
| Qual. | 4.08 ± 0.06; 0.0% | 2.75 ± 0.07; 32.6% | 3.70 ± 0.06; 9.3% | 3.71 ± 0.06; 9.1% | 3.75 ± 0.06; 8.1% | 3.83 ± 0.06; 6.1% |
| Expr. | 4.10 ± 0.05; 0.0% | 2.95 ± 0.07; 28.0% | 3.64 ± 0.05; 11.2% | 3.74 ± 0.05; 8.8% | 3.75 ± 0.05; 8.5% | 3.84 ± 0.05; 6.3% |

**Table 2.** *MOS Evaluation for US English male voice ($\mu\pm95\%$; $100(\mu - \mu_{PCM})/\mu_{PCM}$)*

| MOS | PCM | mod-wrld [9] | mod-lpc [11] | seq2wv (0.2,0.8) | seq2lpc (0.2,0.8) | seq2lpc (0.0,0.8) |
|---|---|---|---|---|---|---|
| Qual. | 4.28 ± 0.05; 0.0% | 2.91 ± 0.07; 32.0% | 3.93 ± 0.05; 8.2% | 4.06 ± 0.05; 5.1% | 4.01 ± 0.05; 6.3% | 4.00 ± 0.05; 6.5% |
| Expr. | 4.30 ± 0.05; 0.0% | 3.09 ± 0.07; 28.1% | 3.86 ± 0.06; 10.2% | 4.07 ± 0.05; 5.3% | 3.99 ± 0.05; 7.2% | 4.00 ± 0.05; 7.0% |

## 6. SUMMARY

In this work we presented the prosody-controllable sequence-to-sequence neural TTS system with LPCNet vocoder backend. The proposed system is capable of real-time speech synthesis on CPU for wide-band (22 kHz) speech synthesis, while preserving the quality of the similar sequence-to-sequence system with Wavenet vocoder backend [7] and significantly improving the quality and the expressiveness of the modular neural TTS with LPCNet backend [11].

---
[2] https://pytorch.org
[3] http://ibm.biz/BdzQrt

# 7. REFERENCES

[1] Wang, Yuxuan, et al. "Tacotron: A fully end-to-end text-to-speech synthesis model." *arXiv preprint arXiv:1703.10135* (2017).

[2] Shen, Jonathan, et al. "Natural tts synthesis by conditioning wavenet on mel spectrogram predictions." 2018 *IEEE International Conference on Acoustics, Speech and Signal Processing (ICASSP)*. IEEE, 2018.

[3] Tamamori, Akira, Tomoki Hayashi, Kazuhiro Kobayashi, Kazuya Takeda, and Tomoki Toda. "Speaker-Dependent WaveNet Vocoder." In *Interspeech*, pp. 1118-1122. 2017.

[4] Valin, Jean-Marc, and Jan Skoglund. "LPCNet: Improving neural speech synthesis through linear prediction." *IEEE International Conference on Acoustics, Speech and Signal Processing (ICASSP)*. IEEE, 2019.

[5] Griffin, Daniel W. and Jae S. Lim. "Signal estimation from modified short-time Fourier transform." *ICASSP (1983)*.

[6] Morise, Masanori, Fumiya Yokomori, and Kenji Ozawa. "WORLD: a vocoder-based high-quality speech synthesis system for real-time applications." *IEICE Trans. on Information and Systems* 99.7 (2016): 1877-1884.

[7] Shechtman, S., Sorin, A. Sequence to Sequence Neural Speech Synthesis with Prosody Modification Capabilities. *Proc. 10th ISCA Speech Synthesis Workshop* (2019), 275-280

[8] Wu, Zhizheng, Oliver Watts, and Simon King. "Merlin: An Open Source Neural Network Speech Synthesis System." *SSW*. 2016.

[9] Kons, Zvi, Slava Shechtman, Alex Sorin, Ron Hoory, Carmel Rabinovitz, and Edmilson Da Silva Morais. "Neural TTS Voice Conversion." In *2018 IEEE Spoken Language Technology Workshop (SLT)*, pp. 290-296. IEEE, 2018.

[10] Van Den Oord, Aäron, et al. "WaveNet: A generative model for raw audio." *SSW* 125 (2016).

[11] Kons, Z., Shechtman, S., Sorin, A., Rabinovitz, C., Hoory, R., "High Quality, Lightweight and Adaptable TTS Using LPCNet." In *Interspeech 2019,* pp. 176-180

[12] Schuster, Mike, and Kuldip K. Paliwal. "Bidirectional recurrent neural networks." *IEEE Transactions on Signal Processing* 45.11 (1997): 2673-2681.

[13] Hochreiter, Sepp, and Jürgen Schmidhuber. "Long short-term memory." *Neural computation* 9.8 (1997): 1735-1780.

[14] Chazan, D., Zibulski, M., Hoory, R. and Cohen, G. "Efficient periodicity extraction based on sine-wave representation and its application to pitch determination of speech signals", in *Proc. Eurospeech 2001*, 2427-2430.